# Dependence of Giant Tunnel Magnetoresistance of Sputtered CoFeB/MgO/CoFeB Magnetic Tunnel Junctions on MgO Barrier Thickness and Annealing Temperature

Jun HAYAKAWA[1,2], Shoji IKEDA[2], Fumihiro MATSUKURA[2], Hiromasa TAKAHASHI[1,2], and Hideo OHNO[2]

*1 Hitachi, Ltd., Advanced Research Laboratory, 1-280, Higashi-koigakubo, Kokubunji-shi, Tokyo 185-8601, Japan*

*2 Laboratory for Nanoelectronics and Spintronics, Research Institute of Electrical Communication, Tohoku University, Katahira 2-1-1, Aoba-ku, Sendai 980-8577, Japan*

We investigated the dependence of giant tunnel magnetoresistance (TMR) on the thickness of an MgO barrier and on the annealing temperature of sputtered CoFeB/MgO/CoFeB magnetic tunnel junctions deposited on $SiO_2$/Si wafers. The resistance-area product exponentially increases with MgO thickness, indicating that the quality of MgO barriers is high in the investigated thickness range of 1.15-2.4 nm. High-resolution transmission electron microscope images show that annealing at 375 °C results in the formation of crystalline CoFeB/MgO/CoFeB structures, even though CoFeB electrodes are amorphous in the as-sputtered state. The TMR ratio increases with annealing temperature and is as high as 260% at room temperature and 403% at 5 K.

There is a great deal of interest in using magnetic tunnel junctions (MTJs), which offer high tunnel magnetoresistance (TMR) ratio[1,2], for non-volatile magnetic random access memories and magnetic read heads for hard disk drives. Most studies to date have been done on MTJs with amorphous aluminum oxide barriers, yielding TMR ratios, defined as the ratio of the resistance change and the resistance of the low resistance state, of 50–70%; the highest reported TMR ratio was 70% in MTJs using conventional 3d ferromagnetic metals such as Fe, Co, and Ni at room temperature (RT).[3] Recent first-principles electronic structure calculations on fully ordered (001) oriented Fe/MgO/Fe MTJs suggest the possibility of TMR ratios of 100 to even 1000% for a sufficiently thick MgO barrier.[4,5] The TMR ratios of body centered cubic Co(001)/MgO(001)/Co(001) and the FeCo(001)/MgO(001)/FeCo(001) systems were predicted to be several times greater than the TMR ratio predicted for the Fe/MgO/Fe system.[6] These giant TMR ratios are thought to be caused by an interfacial spin-dependent electronic state with $\Delta_1$ symmetry at the Fermi energy. Recent experiments at RT have demonstrated giant TMR ratios of up to 180% in single-crystal Fe/MgO/Fe MTJs[7,8], 220% in highly oriented (001) CoFe/MgO/CoFe MTJs[9], and 230% in sputtered CoFeB/MgO/CoFeB MTJs[10]. Djavaprawira et al. obtained results that are especially notable from the technological point of view because they deposited MTJs with standard spin-valve structure on thermal oxidized Si using a conventional sputtering method and then annealed them, obtaining a giant TMR ratio with amorphous CoFeB ferromagnetic electrodes.[10] In this letter, we show how the sputtered CoFeB/MgO/CoFeB MTJs exhibiting giant TMR ratios depend on the thickness of the MgO barrier and on annealing, which has not yet been investigated. We also show that the TMR ratio can be increased by increasing the annealing temperature and can be as high as 260% at RT and 403% at 5 K.

A schematic diagram of the fabricated MTJ devices is shown in Figure 1(a). Multilayer films were deposited using rf magnetron sputtering with a base pressure of $10^{-9}$ Torr. The MTJ films with synthetic pin layers were formed on $SiO_2$/Si substrates. The order of the film layers was as follows, starting from the substrate side; Ta(5) / Ru(50) / Ta(5) / NiFe(5) / MnIr(10) / CoFe(2) / Ru(0.8) / CoFeB(3) / MgO / CoFeB(3) / Ta(5) / Ru(5). The numbers in parentheses indicate the thickness in nm of the layers, and the thickness of the MgO barrier was varied from 1.15 to 2.4 nm. We use CoFe to represent a $Co_{90}Fe_{10}$ alloy and CoFeB to represent a $Co_{40}Fe_{40}B_{20}$ alloy. The MgO layer was formed using an MgO target at a pressure of 1 mTorr in an Ar atmosphere. The MTJs were annealed at 270 to 375 °C for 1 h in a vacuum of $10^{-6}$ Torr under a magnetic field of 4 kOe. All junctions, sized from 0.8 x 0.8 μm$^2$ to 0.8 x 5.6 μm$^2$, were fabricated using a conventional photolithography process. An AFM image of the surface of a Ta/Ru/Ta underlayer (resistivity $\rho$ = 7.5 μΩcm) is shown in Figure 1 (b) and indicates that the surface is smooth with an average roughness ($R_a$) of 0.17 nm. The electrical properties of the MTJs were measured at RT and at 5 K using a four-probe method with a dc bias and a magnetic field of up to 1 kOe. The TMR ratio was measured at 5 K under a field-cooled state (-2 kOe). The current direction was defined as positive when electrons were flowing from the bottom to the top layer.

For an MTJ that was annealed at 375°C for 1h with a 2.0 nm MgO barrier, typical TMR ratios ($\Delta R/R$) versus magnetic field ($H$) loops measured at RT (dashed line) and 5 K (solid line) are shown in Fig. 2 ; the resistance-area product ($RA$) is 3.4 k$\Omega\mu m^2$ at RT and 3.5 k$\Omega\mu m^2$ at 5 K. We found that the TMR ratio is as high as 260% at RT, and reaches 403% at 5 K; these are the highest yet reported TMR ratios of MTJs using conventional 3d ferromagnetic metals and oxide barriers. By using Julliere's formula[11], these TMR ratios are found to correspond to tunneling spin-polarizations of 0.75 and 0.82. The exchange bias field, which gradually decreases with increasing annealing temperature ($T_a$), was 0.35 kOe at RT when the MTJ was annealed at 375°C.

Figure 3 (a) and (b) show $RA$ and $\Delta R/R$ as functions of MgO barrier thickness ($t_{MgO}$). The dotted, dashed, and solid lines show data for MTJs annealed at 270, 325, and 375°C. The exponential increase of $RA$ with increasing $t_{MgO}$ shown in Fig. 3 (a) indicates that pinhole-free barriers with constant barrier height ($\phi$) are formed and maintained at elevated $T_a$. The $\phi$ which is estimated using the Wenzel-Kramer-Brillouin approximation[8,13] is in the range of 0.34–0.38 eV, indicating that it varies little upon annealing. In the ranges from 1.35 to 2.2 nm, the TMR ratio increases dramatically with increasing $T_a$. For example, in a MTJ with a 2.0 nm MgO barrier, the TMR ratio is 100% when $T_a$ = 270°C, 200% when $T_a$ = 325°C, and 260% when $T_a$ = 375°C. However, the ratio does not significantly increase with increasing $T_a$ when $t_{MgO}$ = 2.4 and 1.15 nm. The 110% TMR ratio at $t_{MgO}$ = 1.15 nm, however, is still four times greater than the ratios of aluminum oxide barrier MTJs with similar $RA$s of 30 $\Omega\mu m^{2,12)}$.

The structure of similarly prepared samples without Ru in their underlayers with 240% TMR ratios was explored using high-resolution cross-sectional transmission electron microscopy (HRTEM). The HRTEM images in Fig. 4 show the regions containing the CoFeB(3)/MgO(2)/CoFeB(3) interfaces for samples (a) as-deposited, (b) after annealing at 270°C, and (c) after annealing at 375°C. We can see three features in these images. First, as-deposited CoFeB electrodes have amorphous structures (Fig. 4 (a)), and this structure remains amorphous after annealing at 270°C (Fig. 4 (b)). Second, the MgO barrier has NaCl structure that is highly (001) oriented and has been uniformly deposited on the amorphous CoFeB bottom ferromagnetic layer (Figs. 4 (a) and (b)). Both of these observations are in accordance with a previous report.[10] Third, by annealing at 375°C, where the highest TMR ratio is obtained, full crystallization of the CoFeB ferromagnetic electrodes in body centered cubic structure is observed (Fig. 4 (c)).

The highest TMR ratio together with the crystalline structure observed in the samples annealed at 375°C strongly suggest that the origin of the giant TMR ratios in MgO-based MTJs with amorphous ferromagnetic electrodes resides in the fact that, in addition to the MgO barrier, the ferromagnetic electrodes, at least in the vicinity of the MgO barrier, is of crystalline nature, as Djavaprawire et al.[10] speculated. This may resolve the apparent inconsistency with theory; theoretical studies[4-6] have indicated that the giant TMR is an effect of a particular band structure combination between the crystalline barrier and the electrode material resulting in an effective half-metallic electrode. This combination is not possible when amorphous electrodes are used.

We believe that the crystallization process of CoFeB electrodes is initiated at both MgO/CoFeB interfaces and proceeds to the entire CoFeB layers. This is supported by Fig. 4 (c), which shows smooth MgO/CoFeB interfaces, contrasting with the amorphous Ta/CoFeB upper interface and the rough CoFeB/Ru lower interface.

Current ($I$) is plotted in Figure 5 (a) as a function of bias voltage ($V$) in an MTJ that has a TMR ratio of 260% at RT. The solid line shows the data in a parallel (low-resistance) configuration and the dashed line shows data in an anti-parallel (high-resistance) configuration. Simmons' equation[13] yields $\phi = 0.4$ eV for a barrier thickness d = 2.18 nm using the anti-parallel data, which is in good agreement with values reported previously[8, 9]. The *I-V* curve in the parallel configuration indicates virtually ohmic transport, while the anti-parallel configuration shows typical non-linear tunnel characteristics; note that for MTJs using amorphous aluminum oxide barriers, non-linear *I-V* curves are generally observed in both parallel and anti-parallel configurations. The *I-V* characteristics in the parallel configuration may be due to the matching of the symmetry of the tunneling electronic states predicted by theoretical studies.[4-6]

The normalized TMR ratio(solid line) and output voltage ($\Delta V \sim V_p \times (R_{ap}-R_p)/R_p$, dotted line) are plotted in Fig. 5 (b) as functions of bias voltage. The TMR ratio was found to decrease with increasing bias voltage, and only a slight asymmetry was observed for positive and negative bias voltages compared with that in fully epitaxial Fe/MgO/Fe MTJs.[8] The TMR ratios dropped to half at voltages of about 600 mV at positive bias and about 700 mV at negative bias. This asymmetry may have been caused by the differences in the electronic states at the MgO/CoFeB interfaces. Note that $\Delta V$ at negative bias reaches approximately 450 mV, which is approximately three times that of MTJs with amorphous aluminum oxide barriers.[14]

In conclusion, we reported the dependence of MgO-based MTJs prepared using conventional sputtering on the thickness of MgO barriers and on the annealing temperature. By annealing at 375°C, CoFeB ferromagnetic electrodes separated by highly (001) oriented MgO are fully crystallized in body centered cubic structure. The increase in the TMR ratio due to annealing can be attributed to the formation of highly oriented crystalline CoFeB/MgO/CoFeB MTJs. The obtained TMR ratio in our experiment was as high as 260% at RT and 403% at 5 K. We observed an output voltage of up to 450 mV. The MTJs have characteristic *I-V* responses in parallel configurations, which has not been seen in MTJs with amorphous aluminum oxide barriers.

This work was supported by the IT-program of Research Revolution 2002 (RR2002): "Development of Universal Low-power Spin Memory" from the Ministry of Education, Culture,

Sports, Science and Technology of Japan.

Figure caption

Fig. 1. (a) Schematic drawing of cross section of MTJ, and (b) AFM image of surface of Ta/Ru/Ta underlayer ($\rho$ = 7.5 $\mu\Omega$cm) indicating that surface is smooth with $R_a$ of 0.17 nm.

Fig. 2. Magnetoresistance loops in MTJs at room temperature (dased line) and 5K (solid line). TMR measurement at 5 K was done under field-cooled state (-2 kOe). TMR ratio was as high as 260% at RT and 403% at 5 K.

Fig. 3. (a) *RA* and (b) *ΔR/R* at room temperature as functions of thickness of MgO barriers. Dotted, dashed, and solid lines show data for MTJs annealed at 270, 325, and 375°C.

Fig. 4. Cross-sectional TEM images of CoFeB(3)/MgO(2)/CoFeB(3) interfaces for samples (a) as-deposited, (b) after annealing at 270°C, and (c) after annealing at 375°C. MgO barrier deposited on amorphous CoFeB bottom ferromagnetic layer has NaCl structure that is highly (001) oriented. Amorphous CoFeB ferromagnetic electrodes are crystallized by post annealing at 375°C.

Fig. 5. (a) *I-V* curves of MTJs in parallel configuration (solid line) and in anti-parallel configuration (dashed line). (b) Normalized TMR ratio (solid line) and output voltage *ΔV* (dashed line), defined as $V_p$ x $(R_{ap}-R_p)/R_p$, as functions of bias voltage in MTJs that exhibit TMR ratios of 260%. Bias voltage where TMR ratio was half zero-bias value was 600 mV for positive and 700 mV for negative bias directions. Maximum $\Delta V$ was 450 mV.

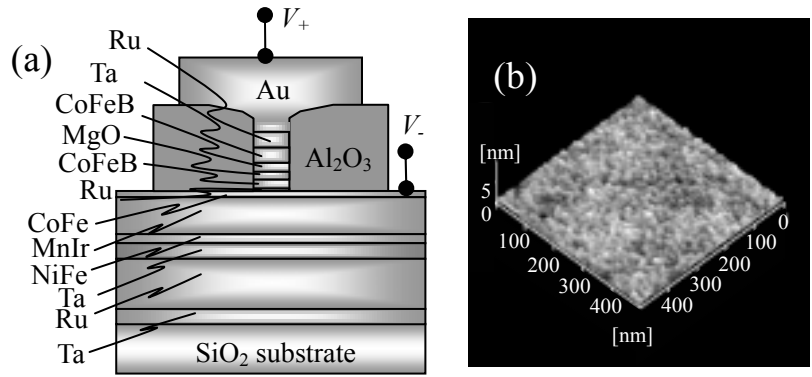

Fig. 1 J. Hayakawa et al

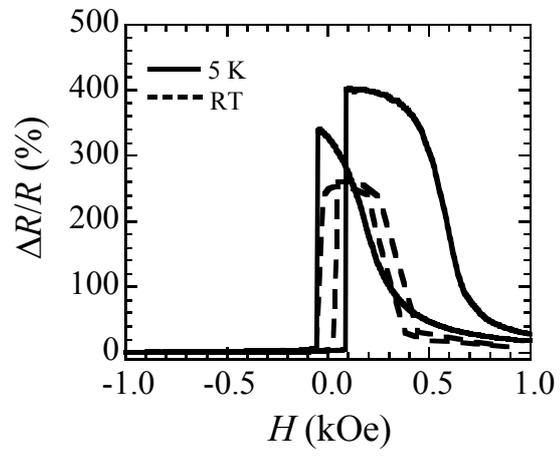

Fig.2　J. Hayakawa et al

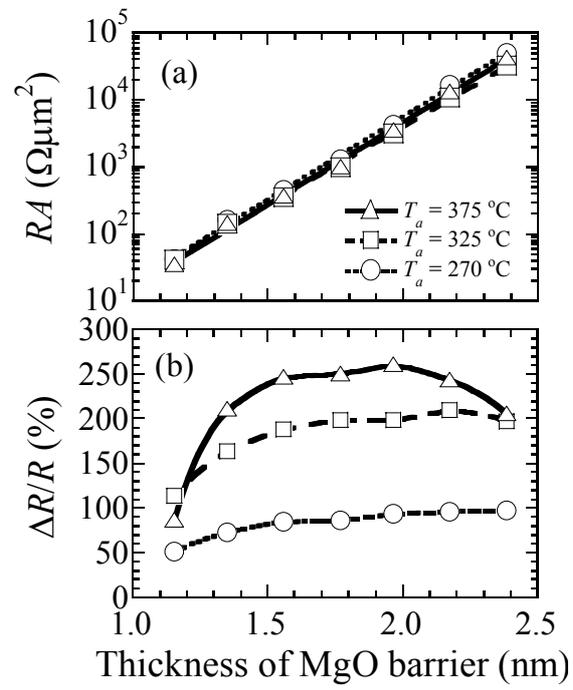

Fig. 3　J. Hayakawa et al

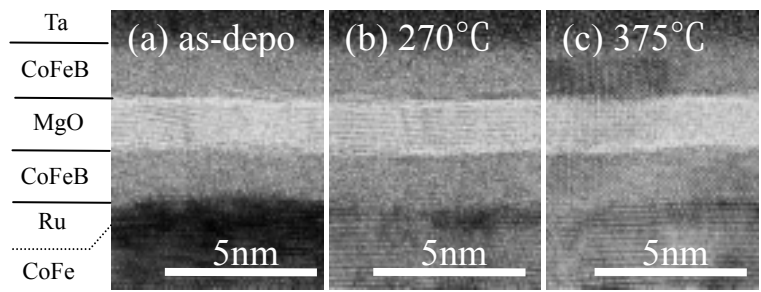

Fig.4　J. Hayakawa et al

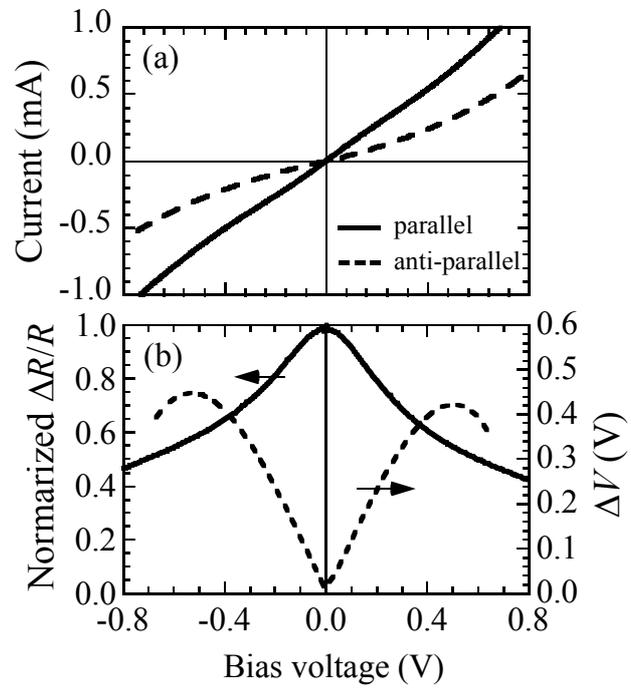

Fig.5  J. Hayakawa et al